\documentclass[notitlepage,a4paper,prd,preprintnumbers,noshowpacs,noshowkeys]
{revtex4}
\usepackage[latin1]{inputenc}
\usepackage{ae,aecompl}
\usepackage{amsmath,amssymb}
\usepackage{mathrsfs}
\usepackage{graphicx}
\usepackage{hyperref}

\providecommand{\aver}[1]{\langle #1 \rangle}
\providecommand{\bs}[1]{\boldsymbol{#1}}
\providecommand{\eq}[1]{\begin{equation} #1 \end{equation}}
\providecommand{\mtrx}[1]{\begin{pmatrix} #1 \end{pmatrix}}
\usepackage{bbm}
\providecommand{\id}{{\mathbbm{1}}} 
\providecommand{\eqali}[1]{\begin{equation}\begin{aligned} #1
    \end{aligned}\end{equation}}
\DeclareMathOperator{\diag}{\mathrm{diag}} 
\providecommand{\ums}[2][1]{\ml{\tfrac{#1}{#2}}} 
\providecommand{\ml}[1]{\mbox{\large $#1$}}
\providecommand{\mss}[1]{\mbox{\scriptsize $#1$}}
\providecommand{\tp}{{\mss{\mathsf{T}}}}
\providecommand{\xlink}[1]
  {\href{http://arxiv.org/abs/#1}{arXiv:#1}}

\providecommand{\tS}{\tilde{S}}
\providecommand{\cp}{\mathsf{CP}}
\providecommand{\hx}{\hat{x}}
\providecommand{\bx}{\mathbf{x}}
\providecommand{\tb}{{\mathrm{TB}}}

\providecommand{\om}{\omega}

\providecommand{\mns}{\mathrm{MNS}}
\providecommand{\lag}{\mathcal{L}}
\providecommand{\eps}{\epsilon}

\begin{document}
\title{
$S_4$ Flavored CP Symmetry for Neutrinos
}
\author{R.~N.~Mohapatra}
\email{rmohapat@umd.edu}
\affiliation{Maryland
Center for Fundamental Physics and Department of Physics,
University of Maryland, College Park, Maryland 20742, USA
}
\author{C.~C.~Nishi}
\email{celso.nishi@ufabc.edu.br}
\affiliation{
Universidade Federal do ABC - UFABC, 
09.210-170,
Santo André, SP, Brasil
}

\preprint{UMD-PP-012-013}
\begin{abstract}
A generalized CP symmetry for leptons is presented where CP transformations 
are part of an $S_4$ symmetry that connects different families. 
We study its implications for lepton mixings in a gauge model realization  of the
idea using type II seesaw for neutrino masses. The model predicts maximal
atmospheric mixing, nonzero $\theta_{13}$ and maximal Dirac phase
$\delta_{D}=\pm\frac{\pi}{2}$.
\end{abstract}
\maketitle
\section{Introduction}
\label{sec:intro} 

The recent measurement of the leptonic mixing angle $\theta_{13}$ in experiments
searching for the  oscillations of electron antineutrinos  emitted from
reactors~\cite{dayabay} and from the accelerator based experiments with muon
neutrino beam~\cite{t2k} has generated considerable excitement in the field of
neutrino physics. Taken together with already measured solar angle $\theta_{12}$ and
atmospheric angle $\theta_{23}$, this almost completes the CP-conserving part of the
lepton
mixing matrix, under the assumption that there are no sterile neutrinos.  This 
narrows the focus of the field to three  remaining unknowns of neutrino flavor
physics: (i) Dirac versus Majorana nature of the neutrino masses, (ii) mass
hierarchy among them---namely, normal versus inverted---and (iii) leptonic
CP-violating phases. The last item has two parts to it: Dirac phase, which is
analogous to the CKM phase in quark sector, and Majorana phases, which are exclusive
to the neutrino sector for Majorana neutrinos. The former can be measured in
oscillation experiments whereas the latter may play a role in neutrinoless double
beta decay searches. All these phases may play a role in understanding the origin of
matter.

On the theoretical side, despite such a vast amount of information, the nature of BSM physics responsible for neutrino flavor properties  remains largely unknown and is the subject of extensive investigation. There are two generic approaches: one based on symmetries in the lepton sector leaving the quarks aside and a second one based on grand unified theories where both quarks and leptons are considered together. 

The quark-lepton unified grand-unification-based approach not only provides a very
natural embedding of the seesaw mechanism to explain small neutrino masses but also,
in a very economical class of renormalizable SO(10) models, turns out to be very
predictive. Indeed,  the recently measured value of $\theta_{13}$ agrees with
predictions made for this parameter in a minimal model of this type in
2003--2005\,\cite{theta}.   While this agreement is impressive, until there is some
other evidence directly connecting the grand unification properties to seesaw
physics (e.g., B-L violation as in \cite{bm1}), one cannot test the GUT seesaw
approach. 

The  symmetry approach, on the other hand, derives its appeal from the fact that two
of the observed neutrino mixing angles $\theta_{23}$ (atmospheric) and $\theta_{12}$
(solar)  are close to  values that look like group theoretical numbers, and find
easy explanation in terms of simple discrete family symmetry-based models. For
example, the observed near-maximal atmospheric mixing can
be easily understood if, in a basis where charged lepton masses are diagonal (to be
called ``flavor basis'' from here on), the Majorana neutrino mass matrix satisfies
the $Z_2$ $\mu$-$\tau$ symmetry\,\cite{mutau}. The simple versions of this symmetry,
however, predict vanishing $\theta_{13}$, a result which is contradicted by recent
reactor\,\cite{dayabay} and accelerator experiments\,\cite{t2k}. There is vast
literature on the corrections to $\mu$-$\tau$ symmetry that  come either from
allowing general forms for the charged lepton matrix or from changing the neutrino
mass matrix itself or combining simple $\mu$-$\tau$ symmetry with simultaneous CP
conjugation\,\cite{mutau:ref,grimus:gcpnu}. 
All these cases lead to nonzero $\theta_{13}$. Many such models
are also now ruled out since they predict values of $\theta_{13}$ much smaller than
the measured value. 
If, in addition to maximal atmospheric mixing, we consider the value of the solar
angle $\tan \theta_{12}\simeq\frac{1}{\sqrt{2}}$, we obtain the so-called
tri-bimaximal mixing\,\cite{tbm}, and it suggests more complicated groups such as
$Z_2\times Z_2$\,\cite{z2xz2} or $S_3$\,\cite{yu} or $A_4$\,\cite{ma}, but some of
them also imply that $\theta_{13}$ is zero or small after charged lepton corrections
are taken into account and are not anymore phenomenologically viable. Thus the
measurement of $\theta_{13}$ has had a great impact on neutrino model building.

The discovery of large $\theta_{13}$, however, does not rule out the generic
symmetry approach and many examples have been discussed where new symmetries do
allow for a large nonzero $\theta_{13}$~\cite{joshipura,ishimori,alt,others}. We
discuss one such approach in this paper which not only has the virtue of allowing
large $\theta_{13}$ but also predicts all the leptonic CP phases. The approach is
somewhat different from many papers in the sense that we use a generalized
definition of CP transformation among leptons\,\cite{grimus:fcnc} embedded in an
$S_4$ lepton family symmetry. We will call this new symmetry $\tilde{S}_4$ symmetry.
We present a gauge model for leptons invariant under this symmetry which not only
accommodates a large $\theta_{13}$ but also predicts a maximal $\theta_{23}$ and
maximal Dirac CP phase, i.e., $\delta_D=\pm\frac{\pi}{2}$. The maximal $\theta_{23}$
is still consistent with the latest global analysis\,\cite{fogli.12,valle.12},
although there are indications that it may be smaller\,\cite{fogli.12}.

This paper is organized as follows: In Sec.\,\ref{sec:model}, we present the
$\tilde{S}_4$ model and the generalized CP transformation used in it; in
Sec.\,\ref{sec:prediction} we present the various predictions of the model. In
Sec.\,\ref{sec:conclusion}, we give some comments and conclude with a summary of the
results. In an appendix, we discuss the representations of the $\tS_4$ symmetry that
we use in the paper.

\section{Model}
\label{sec:model}

Our model is based on the standard model gauge group $SU(2)_L\otimes U(1)_Y$ with
the usual assignment for leptons. Namely, the left-handed leptons $L_i$ transform as
$SU(2)_L$ doublets with $Y=-1$ and the right-handed charged leptons $l_i$
($=l_{iR}$) transform as singlets with $Y=-2$. The charged leptons gain masses
through the Yukawa interactions with three Higgs doublets $\phi_i \sim (2,1)$,
$i=1,2,3$. Neutrino masses and mixing are generated through type II seesaw
mechanism\,\cite{type2} which requires the introduction of $Y=2$ $SU(2)_L$ triplets.
In order to implement the symmetry in our model, we introduce  four SM triplets,
$\Delta_0$ and $\Delta_i\sim (3,2)$, $i=1,2,3$,  whose neutral members  acquire
small vacuum expectation values (vevs), induced by trilinear couplings of the form
$\phi\phi\Delta^\dagger$. We assume only three families of leptons and no singlet
sterile neutrinos.

We assume the theory to be invariant under a flavor symmetry acting in the horizontal
space of the replicated fields.
The chosen group is isomorphic to $S_4$, but will \textit{contain} generalized CP
transformations (GCP) defined below; we denote this group by $\tS_4$. Note that
the group $S_4$ has been pointed out as the group for tri-bimaximal
mixing\,\cite{lam:s4}, although some subgroup of it may turn out to be just
accidental\,\cite{z2xz2,hernandez}.
The action of $\tS_4$ on complex fields will be nontrivial.
It is constructed as a subgroup of $S_4\otimes \aver{\cp}$ as
follows. We remind the reader that $S_4$ has generators $S$ and $T$ which satisfy
the properties $S^4=T^3=\id$ and $ST^2S=T$.

Let us consider the (faithful) three-dimensional representation $\bs{3}$ of
$S_4$ generated by\,\cite{hagedorn}
\eq{
\label{S4:rep3}
\bs{3}: \quad
S=\mtrx{-1 & 0 & 0\cr 0 & 0 & -1 \cr 0 & 1 & 0}\,,\quad
T=\mtrx{0 & 0 & 1\cr 1 & 0 & 0 \cr 0 & 1 & 0}\,.
}
For complex fields, we can adjoin the usual CP transformation, denoted by the
operator $\cp$, to obtain $S_4\otimes \aver{\cp}$. Note that $S_4$
transformations and the CP transformation commute because all representations of
$S_4$ are real.
We then extract the subgroup of $S_4\otimes\aver{\cp}$ generated by
\eq{
\label{tS4:rep3}
\bs{3}: \quad
\tS=\mtrx{-1 & 0 & 0\cr 0 & 0 & -1 \cr 0 & 1 & 0}\!\cdot\cp\,,\quad
T=\mtrx{0 & 0 & 1\cr 1 & 0 & 0 \cr 0 & 1 & 0}\,.
}
Notice the charge conjugation part in $\tS$ is trivial for real fields.
This group is isomorphic to $S_4$ after we factor the subgroup generated by
$\cp^2=-\id$ for fermions. Such a factor group is $\tS_4$.
We keep the notation $\bs{3}$ for the representation generated by \eqref{tS4:rep3}.
The other representations of $\tS_4$ should be constructed in a similar manner from
the representations $\bs{3}',\bs{2},\bs{1}',\bs{1}$ of $S_4$.
It is important to point out that $\tS$ is a nontrivial GCP transformation that
does not reduce to the usual CP transformation by basis
change\,\cite{grimus:fcnc}.

Let us list the irreducible representations (irreps) of $\tS_4$, constructed from
the irreps $\bs{1},\bs{1}',\bs{2},\bs{3},\bs{3}'$ of $S_4$.
They are led to peculiar representations of $\tS_4$ when complex fields are
considered: the real irreps $\bs{1}$ and $\bs{1}'$ ($\bs{3}$ and
$\bs{3}'$) are interwoven in one equivalent (complex) representation $\bs{1}$
($\bs{3}$) whereas $\bs{2}$ splits into two inequivalent complex
one-dimensional representations which we denote by $\bs{1}_\om$ and
$\bs{1}_{\om^2}$; see Appendix \ref{ap:S4cp:2} for an explanation. They are
quite similar to the representations $\bs{1}',\bs{1}''$ of $A_4$.

We assign the representations of $\tS_4$ as follows:
\eq{
L_i\sim\bs{3},\quad
l_1\sim \bs{1},~
l_2\sim \bs{1}_\om,~
l_3\sim \bs{1}_{\om^2},\quad
\phi_i\sim\bs{3}\,.
}
The fields assigned to the triplet representation \eqref{tS4:rep3} transform
explicitly as
\begin{equation}
\begin{aligned}
L_i(x)  &\stackrel{\tS}{\longrightarrow} S_{ij}CL_j^*(\hx) \,,&
L_i(x)  &\stackrel{T}{\longrightarrow} T_{ij}L_j(x)\,; 
\cr
\phi_i(x) &\stackrel{\tS}{\longrightarrow} S_{ij}\phi_j^*(\hx) \,,&
\phi_i(x) &\stackrel{T}{\longrightarrow} T_{ij}\phi_j(x)\,,
\end{aligned}
\end{equation}
where $\hx=(x_0,-\bx)$ for $x=(x_0,\bx)$ arises because of space inversion and $C$
is the charge conjugation matrix.
On the other hand, the right-handed lepton fields transform as
\begin{equation}
\begin{aligned}
l_1(x)  &\stackrel{\tS}{\longrightarrow} Cl_1^*(\hx) \,,&
l_1(x)  &\stackrel{T}{\longrightarrow} l_1(x)\,; 
\cr
l_2(x)  &\stackrel{\tS}{\longrightarrow} Cl_2^*(\hx) \,,&
l_2(x)  &\stackrel{T}{\longrightarrow} \om\,l_2(x)\,; 
\cr
l_3(x)  &\stackrel{\tS}{\longrightarrow} Cl_3^*(\hx) \,,&
l_3(x)  &\stackrel{T}{\longrightarrow} \om^2 l_3(x)\,.
\end{aligned}
\end{equation}

The Yukawa interactions for charged leptons invariant under these transformations is
given by 
\eqali{
\label{Y:l}
-\lag_Y^l=&~y_1(\bar{L}_1\phi_1+\bar{L}_2\phi_2+\bar{L}_3\phi_3)l_1
  +y_2(\bar{L}_1\phi_1+\om^2\bar{L}_2\phi_2+\om\bar{L}_3\phi_3)l_2
\cr
  & + y_3(\bar{L}_1\phi_1+\om\bar{L}_2\phi_2+\om^2\bar{L}_3\phi_3)l_3
  + h.c.,
}
with the important restriction that all couplings $y_i$ are \textit{real} due to
invariance by $\tS$.

When the neutral parts of the Higgs doublets acquire the vevs
\eq{
\label{vev:phi}
\aver{\phi_i}=\frac{v}{\sqrt{3}}(1,1,1)\,,
}
the Lagrangian \eqref{Y:l} gives rise to the charged lepton mass matrix
\eq{
\label{Ml}
M_l=\frac{1}{\sqrt{3}}
\begin{pmatrix}
1 & 1 & 1\cr1 & \om^2 & \om\cr 1 & \om & \om^2
\end{pmatrix}
\diag(m_e,m_\mu,m_\tau)\,.
}
The correspondence is $(m_e,m_\mu,m_\tau)=v(y_1,y_2,y_3)$ and we identify $U_\om^*$
in \eqref{Ml} by defining
\eq{
U_\om\equiv
\frac{1}{\sqrt{3}}\begin{pmatrix}
1 & 1 & 1\cr1 & \om & \om^2\cr 1 & \om^2 & \om
\end{pmatrix}\,.
}
We can see that $M_lM_l^\dag$ has circulant form\,\cite{tbm} and it is invariant by
$T$ and any transposition of family indices composed with complex conjugation (CP
transformation), i.e., an $\tS_3$ subgroup of $\tS_4$. The matrix \eqref{Ml} is
identical to the one obtained in $A_4$ models. The potential for $\phi_i$ is in fact
the same as the general $A_4$ invariant potential\,\cite{ma}, implying that $A_4$
invariance leads automatically to $\tS_4$ invariance for the potential of three
Higgs doublets. For that potential, it has been shown that \eqref{vev:phi} is a
possible minimum\,\cite{merlo:a4}.

To generate neutrino masses, we introduce four Higgs triplets transforming under
$\tS_4$ as
\eq{
\Delta_0\sim\bs{1},\quad
\Delta_i\sim\bs{3}\,.
}
The $\tS_4$-invariant Lagrangian is then
\eq{
\label{L:nu}
-\lag^\nu=\ums{2}f_0\overline{L_i^c}\eps\Delta_0L_i
  +f_1\big(\overline{L_2^c}\eps\Delta_1L_3
    +\overline{L_3^c}\eps\Delta_2L_1
    +\overline{L_1^c}\eps\Delta_3L_2\big)
  +h.c.,
}
where $f_0,f_1$ are also real due to $\tS$.

Given the large vev hierarchy, we can assume the potential allows arbitrary
vevs for the neutral components of $\Delta_0,\Delta_i$,
\eq{
\label{vev:Delta}
\aver{\Delta_0^{(0)}}=u_0\,,\quad
\aver{\Delta_i^{(0)}}=u_i\,.
}
The Lagrangian \eqref{L:nu} then induces the neutrino mass matrix
\eq{
\label{Mnu:adef}
M_\nu=
\begin{pmatrix}
a & f & e\cr f & a & d\cr e & d & a
\end{pmatrix}\,,
}
where $a=f_0u_0$, $d=f_1u_1,e=f_1u_2,f=f_1u_3$.
Notice the tri-bimaximal limit corresponds to $e=f=0$\,\cite{tbm}.
For real $a,d$ and complex $e=f^*$, the symmetry corresponding to $23$-transposition
and complex conjugation (corresponding to an element of $\tS_4$) would remain
unbroken in the theory as symmetries of $M_\nu$ and $M_lM_l^\dag$. This would lead
to CP invariance  and nonzero $\theta_{13}$.
In contrast, if $e=f$ we would obtain $\theta_{13}=0$.
In our case, CP violation and $\theta_{13}\neq 0$ are allowed because there is no
relation between $e$ and $f$.

If we assume the vevs \eqref{vev:Delta} are real, the neutrino mass matrix, in the
basis where the charged-lepton mass matrix is diagonal, is given by
\eq{
\label{Mnu:lD}
U_\om^\dag M_\nu U_\om^*=
\begin{pmatrix}
x & z & z^* \cr
z & -2z^* & y \cr
z^*  & y  & -2z
\end{pmatrix}\,,
}
where $x,y$ are real while $z$ is in general complex; they are independent
combinations of the four parameters $a,d,e,f$ in \eqref{Mnu:adef}.
This matrix has the same form as in \cite{babu.valle}, invariant by $\mu\tau$
exchange composed with complex conjugation 
(called $\mu\tau$-reflection in \cite{mutau:ref}), 
with additional constraints so that it
depends only on four real parameters.
It has been shown that this form of the mass matrix leads to maximal $\theta_{23}$
and maximal CP violation\,\cite{grimus:gcpnu}, with $\theta_{13}\neq 0$.

The lepton mixing matrix $V_{\mns}$ will be the matrix that diagonalizes
\eqref{Mnu:lD}. It is experimentally known that $V_{\mns}$ is close to the
tri-bimaximal mixing matrix
\eq{
\label{def:UTB}
U_\tb=
\begin{pmatrix}
\sqrt{\frac{2}{3}} & \ums{\sqrt{3}} & 0\cr
-\ums{\sqrt{6}} & \ums{\sqrt{3}} & -\ums{\sqrt{2}}\cr
-\ums{\sqrt{6}} & \ums{\sqrt{3}} & \ums{\sqrt{2}}
\end{pmatrix}\,.
}
Therefore we parametrize
\eq{
\label{MNS}
V_\mns=U_\tb\diag(1,1,i)U_\eps\,,
}
where $U_\eps$ is the matrix that diagonalizes 
\eq{
\label{Mnu'}
M_\nu'=
U^\tp M_\nu U=
\begin{pmatrix}
 a+d & b & 0 \cr
 b & a & c \cr
 0 & c & a-d
\end{pmatrix}\,,
}
for $U=U_\om^*U_\tb\diag(1,1,i)$, $b=\frac{e+f}{\sqrt{2}}$
and $c=\frac{e-f}{\sqrt{2}}$, with $a,d,e,f$ being the original real parameters in
\eqref{Mnu:adef}. 
This mass matrix has the same form as in the $A_4$ model of \cite{ishimori} but our 
definition differs from \cite{ishimori} in that \eqref{MNS} includes an
additional factor of $i$ in $\diag(1,1,i)$.
Therefore, our case corresponds to taking $c$ purely imaginary in \cite{ishimori}. 
However, this case was not considered there because it was focused on nonmaximal
$\theta_{23}$ and both real and imaginary parts were allowed to vary.
In contrast, real $a,d,e,f$ in the matrix \eqref{Mnu:adef} and,
consequently, maximal $\theta_{23}$, are natural consequences of our choice of
symmetry.

We can assume $c>0$ and consider the case $c<0$ by replacing $i$ with $-i$ in
\eqref{MNS}.
This means that the sign of the Dirac phase $\delta_D=\pm\frac{\pi}{2}$ is not
predicted in this model.
Note that $c$ controls $\theta_{13}\neq 0$ (and CP violation)
and therefore it must be nonzero. 

The limit $b,c\to 0$ leads to the tri-bimaximal form as $U_\eps=\id$.
As $c\neq 0$ to guarantee $\theta_{13}\neq 0$, $U_\eps$ should deviate from
the identity. That means $M_\nu'$ must be nearly diagonal, i.e., $|b|,|c|\ll
|a|,|d|$.
Having four parameters to describe 9 quantities, we have 5 predictions, some of
which are independent of the values of $a,b,c,d$. This is a consequence of the 
specific  form of the mass matrix \eqref{Mnu:lD}, i.e., maximal $\theta_{23}$ and
maximal CP violation\,\cite{grimus:gcpnu}.
In our specific model, the Majorana phases are also fixed: one is maximal and the
other is zero. Only  normal mass hierarchy for neutrinos is allowed.
The remaining 5 physical quantities---two angles $\theta_{12},\theta_{13}$ and
three neutrino masses  $m_1,m_2,m_3$---are correlated as they depend only on four
parameters $a,b,c,d$ as discussed in the next section.

A few comments are in order before we proceed to present the detailed numerical analysis of the model. 

\begin{itemize}

\item It is worth noting that in our model the lightest two neutrino eigenstates are
almost degenerate in mass and are about a factor of three lighter than the third
eigenstate unlike most normal hierarchy models where $m_2/m_3 \sim 0.2$ or so. 

\item The Higgs potential for doublet fields in our model is the same as in
the $A_4$ models discussed in \cite{merlo:a4} and it is easy to see from there that
there is a range of parameters in the scalar self-couplings where the
vacuum alignment of the doublet fields in our model is justified.

\end{itemize}

\section{Predictions}
\label{sec:prediction}

In the limit $b,c\to 0$, the neutrino masses, i.e., the absolute values
of the eigenvalues of \eqref{Mnu'}, are given by
\eq{
\label{mi:||:0}
m_1=|a+d|\,,\quad m_2=|a|\,,\quad m_3=|a-d|\,.
}
We can choose $a>0$. From $\Delta m^2_{12}=m_2^2-m^2_1>0$, we can see that $d<0$,
hence normal hierarchy is the only possibility.
The experimental information $\Delta m^2_{23}=m^2_3-m^2_2\gg \Delta m^2_{12}$
allows us to eliminate the modulus symbols in \eqref{mi:||:0} as 
\eq{
\label{mi:0}
m_1=|d|-a\,,\quad m_2=a\,,\quad m_3=a+|d|\,.
}
We then arrive at the sum rule
\eq{
\label{sum}
m_3-2m_2-m_1=0\,,
}
which commonly arises in models with discrete flavor symmetries\,\cite{sumrule}.
The difference here is that the sum rule \eqref{sum} applies to the neutrino
masses themselves without additional Majorana phases or signs.

When we allow $b,c\neq 0$, the sum rule \eqref{sum} is still exactly
satisfied provided that $b=\pm c$. This can be seen from the eigenvalues of
\eq{
\delta M_\nu'\equiv M_\nu'-a\id_3\,,
}
which has characteristic equation
\eq{
\label{det=0}
-\det(\delta M_\nu'-\lambda\id_3)=
\lambda^3-(d^2+b^2+c^2)\lambda-d(b^2-c^2)=0\,.
}
The eigenvalues of $M_\nu'$ can be obtained from the roots of \eqref{det=0} by
adding $a$.

For general $b$ and $c$ the sum rule \eqref{sum} is only valid approximately.
The violation of the sum rule is quantified by
\eq{
\label{def:eps}
\eps_b\equiv -\frac{b}{d}\,,\quad
\eps_c\equiv -\frac{c}{d}\,,
}
which controls the deviation of the PMNS matrix \eqref{MNS} from the tri-bimaximal
mixing \eqref{def:UTB}.
The characteristic equation \eqref{det=0} shows that neutrino masses depend, apart
from $a$, only on two combinations of $d,c,b$ which can be chosen as
\eq{
d'\equiv |d|\sqrt{1+\eps_b^2+\eps_c^2}\,,\quad
\delta \equiv \frac{\eps_c^2-\eps_b^2}{[1+\eps_b^2+\eps_c^2]^{3/2}}\,.
}
We can see $\delta$ quantifies the violation of the sum rule.

We can seek approximate roots to \eqref{det=0} for $|\delta|\ll 1$, which leads to
\eqali{
\label{mnu}
-m_1&= a-d'(1-\ums{2}\delta) \,,\cr
m_2&= a-d'\delta \,,\cr
m_3&= a+d'(1+\ums{2}\delta) \,.
}
The result is valid up to terms of order $\delta^2$ (order $\eps^4$) multiplied 
by $d'$.
These relations can be inverted to write $a,d',\delta$ in terms of the masses.
In particular, the deviation of the sum rule is given by
\eq{
\label{sum:delta}
m_3-2m_2-m_1=\frac{3}{2}\delta\,(m_3+m_1)\,.
}
The knowledge of $\Delta m^2_{23}$ and $\Delta m^2_{12}$ determines the parameters
$a,d'$ in terms of $\delta$.
In turn, $\delta$ depends on $\eps_b$ and $\eps_c$, which affect $\theta_{12}$ and
$\theta_{13}$.

To see how the mixing angles $\theta_{12}$ and $\theta_{13}$ are affected by
$\eps_b,\eps_c$, we can perform an analysis similar to \cite{ishimori}, with the
difference that we have real matrices in our case.
The matrix $U_\eps$ quantifies the deviations of the lepton mixing matrix from the
tri-bimaximal form.
For $M_\nu'$, given the eigenvalues $(-m_1,m_2,m_3)$ in \eqref{mnu}, we can
calculate
the eigenvectors which make up $U_\eps$.
The first approximation leads to
\eq{
U_\eps\approx
\begin{pmatrix}
1 & \eps_b & 0 \cr
-\eps_b & 1 & \eps_c \cr
0 & -\eps_c & 1 
\end{pmatrix}\,,
}
where the real parameters $\eps_b,\eps_c$ were given in \eqref{def:eps}.
Notice $\eps_c>0$ for $c>0$ because $d<0$.

To first order, $\theta_{13}$ depends only on $\eps_c$, while $\theta_{12}$ depends
on $\eps_b$, as
\eqali{
\sin^2\!\theta_{13}&\approx\ums{3}\eps_c^2\,,\cr
\sin^2\!\theta_{12}&\approx\frac{1}{3}+\frac{2\sqrt{2}}{3}\eps_{b}\,.
}
We can then approximate
\eq{
\delta\approx 3\,s^2_{13}-\frac{9}{8}\Big(s^2_{12}-\frac{1}{3}\Big)^2\,,
}
where $s^2_{13}\equiv \sin^2\!\theta_{13}$ and $s^2_{12}\equiv \sin^2\!\theta_{12}$
as usual.
This is the amount of deviation for the sum rule \eqref{sum:delta}.
We can see that the data\,\cite{valle.12} are compatible with $\eps_b\approx 0$.

Given the experimentally known values of $\Delta m^2_{12},\Delta
m^2_{23},\theta_{12},\theta_{23},\theta_{13}$, we can determine
the values of the neutrino masses
\eq{
m_1\approx 13.3\,{\rm meV}\,,\quad
m_2\approx 15.9\,{\rm meV}\,,\quad
m_3\approx 52.1\,{\rm meV}\,.
}
We have used the best-fit values of Ref.\,\cite{valle.12}.
A more precise numerical study reveals that
\eq{
11.8\,\mathrm{meV}\le m_1 \le 13.6\,\mathrm{meV}\,,
}
when the $1\sigma$ range for the observables is allowed\,\cite{valle.12}; see
figures below.

Analogously, we can see the deviation for the sum rule is small as
$\ums[3]{2}\delta\sim 0.1$.
In fact, our numerical study quantifies the deviation as
\eq{
\frac{m_3-2m_2-m_1}{m_3+m_1}=11\% \text{ to } 15\%\,,
}
at the $1\sigma$ interval.

\begin{figure}[h]
\centering
\includegraphics[scale=0.3,angle=90]{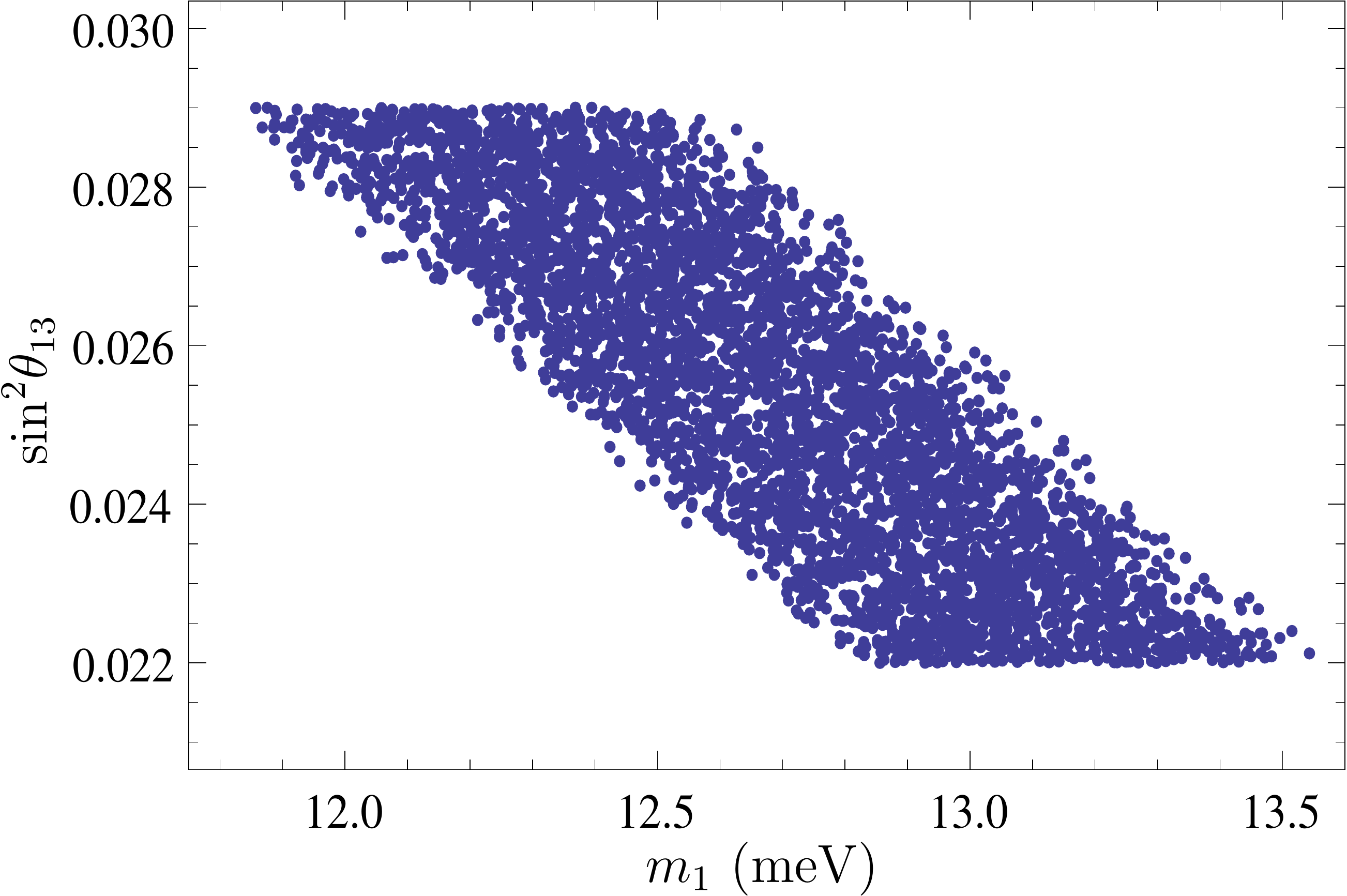}
  \caption{Variation of $\theta_{13}$ as a function of the lightest neutrino mass.  This scatter plot was generated using $\epsilon_c > 0$.
There are points with $\epsilon_c<0 $ as well, corresponding to
flipping the sign of the Dirac phase. However they do not introduce any
perceptible change.}
\label{sym}
\end{figure}

\begin{figure}[h]
\centering
  \includegraphics[scale=0.3,angle=90]{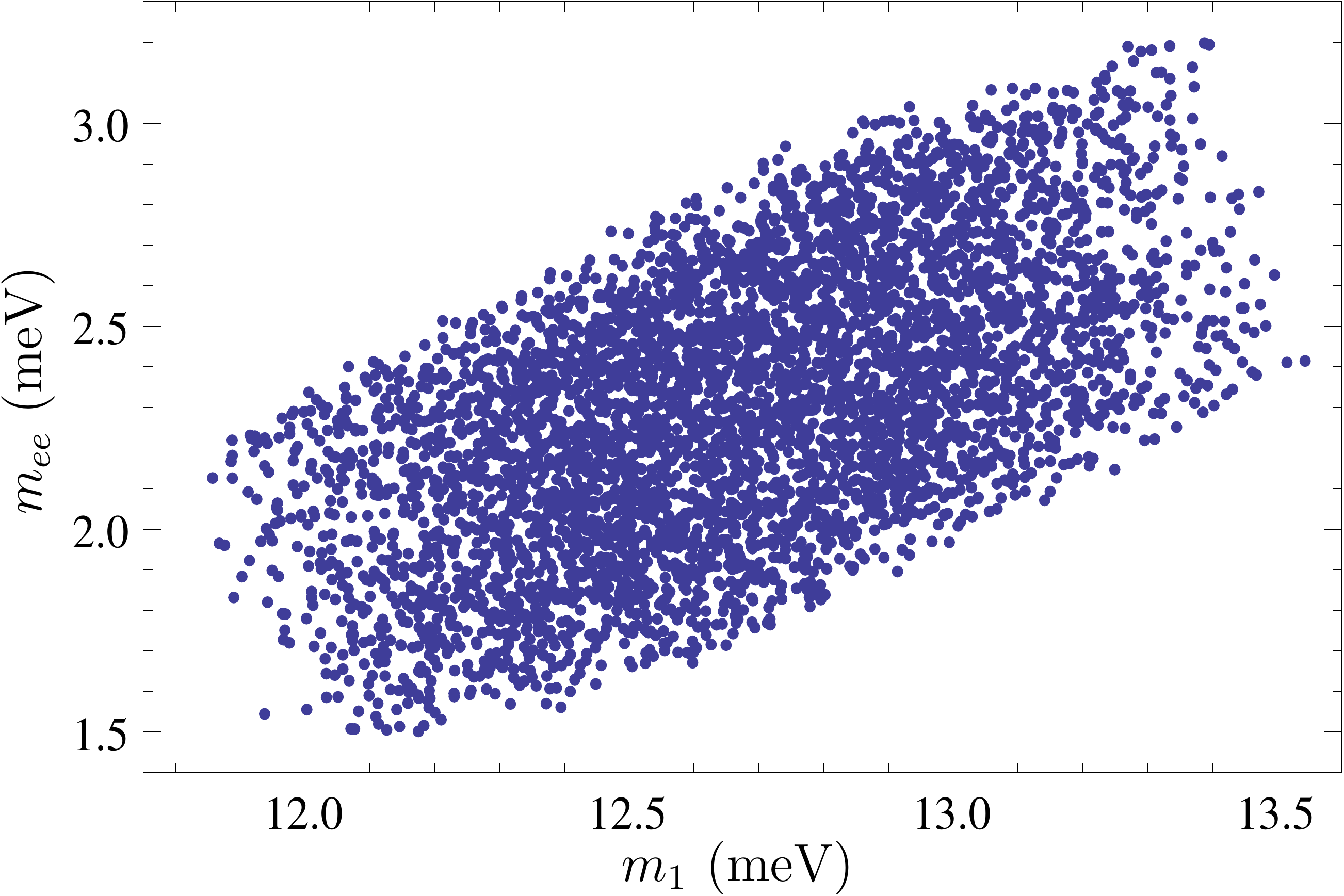}
  \caption{The effective neutrino mass measured in neutrinoless double beta
decay as a function of the lightest neutrino mass. As in Fig.\,\ref{sym}, 
we have chosen $\epsilon_c > 0$ here.}
\label{sym2}
\end{figure}

The remaining numerical study is summarized in two figures.
In Fig.\,\ref{sym} we display the range of $\sin^2\theta_{13}$ against the lightest
neutrino mass. In Fig.\,\ref{sym2}, we display the effective light  neutrino
contribution $m_{ee}$ to neutrinoless double beta decay. Even though the two
light neutrinos are quite degenerate in mass and have masses near 12 meV, due to
Majorana phase, the effective mass is at most 3 meV. 
For both graphics the points are generated numerically without the analytic
approximations employed in the previous analyses.
We only collect the points compatible with the observables within
1-$\sigma$ as shown in Ref.\,\cite{valle.12}.

\section{Conclusions} 
\label{sec:conclusion}

We have presented a model for leptons based on generalized CP symmetries which
transform one family to another, generating the non-abelian $S_4$ symmetry when
supplemented by some permutations of families.
This flavor symmetry, denoted by $\tS_4$, represents a new implementation
of the $S_4$ symmetry where generalized CP symmetries are part of the group. 
This implementation shares some common features with the widely used group $A_4$.
For example, $\tS_4$ also possesses three inequivalent one-dimensional 
representations that is similar to $A_4$  in model building.
The presence of CP transformations, however, further restricts the parameters of
the Lagrangian to be real. 
The restrictions imposed by the generalized CP transformations are such that, with
the addition of another Higgs doublet, we could have easily built another
variant of the model where left-handed and right-handed leptons are assigned to the
same representation $\bs{3}$ of $\tS_4$.
This could help us to embed this type of model in more symmetric theories such as
left-right models.
Therefore, this class of symmetries containing generalized CP
transformations presents interesting features which can be further explored for
flavor model building.

Our specific model predicts maximal atmospheric mixing angle and accommodates
the observed $\theta_{13}$ without any cancellation among the model parameters; it
predicts normal hierarchy and maximal Dirac phase of $\pm90^\circ$ in the leptonic
sector and should be testable in near-future long-baseline neutrino oscillation
experiments. An important feature of the model is that the two light neutrino mass
eigenstates are nearly degenerate in mass. Although the individual light eigenstates
are ``heavy,'' i.e., near 12 meV or so, due to maximal Majorana phase their net
contribution to neutrinoless double beta decay amplitude is very small.
The model also predicts an approximate sum rule relation valid for the three
neutrino masses, without any Majorana phase or sign. The validity of the
approximate sum rule is around 12\%.

\acknowledgements

The work of  R.N.M. has been supported by National
Science Foundation Grant No.\ PHY-0968854.
The work of C.C.N. was partially supported by Brazilian CNPq and Fapesp.
C.C.N. also thanks the \emph{Maryland Center for Fundamental Physics} for their
hospitality during the development of this work. 

\appendix 
\section{Other representations of $\tS_4$}
\label{ap:S4cp:2}

We show here how to obtain the representations $\bs{1}_\om$ and
$\bs{1}_{\om^2}$ of $\tS_4$ from the irreducible representation (irrep) $\bs{2}$ of
$S_4$. The irreps of $\tS_4$ are constructed from the irreps of $S_4$ by the
procedure explained in Sec.\,\ref{sec:model} for the representation $\bs{3}$:
extract the subgroup of $S_4\otimes\aver{\cp}$ generated by $\tS$ and $T$ in
\eqref{tS4:rep3}, instead of $S,T,\cp$ that generate $S_4\otimes\aver{\cp}$.

Let us first explain why $\bs{1}$ and $\bs{1}'$ of $S_4$ generate the same
representation $\bs{1}$ of $\tS_4$. 
We know that, for $S_4$, $\bs{1}$ is trivial but $\bs{1}'$ changes sign by $S$. 
If we follow the recipe and construct the representation of $\tS_4$ corresponding to
$\bs{1}$ and $\bs{1}'$ we would obtain
\eqali{
\label{tS4:11'}
\bs{1}:&~ \tS=S\cdot\cp \to 1\cdot\cp\,,\quad
T\to 1\,,
\cr
\bs{1}':&~ \tS=S\cdot\cp \to (-1)\cdot\cp\,,\quad
T\to 1\,.
}
We are using the generators \eqref{tS4:rep3} of $\bs{3}$ of $\tS_4$ as the group
elements themselves, given that the representation is faithful.
The CP transformation denoted by $\cp$ acts as usual. A fermion field $\psi(x)$ and
a complex scalar field $\phi(x)$ transform as
\eqali{
\label{cp:def}
\psi(x)&\stackrel{\cp}{\longrightarrow} C\psi^*(\hx)\,\cr
\phi(x)&\stackrel{\cp}{\longrightarrow} \phi^*(\hx)\,.
}
Therefore, the representations \eqref{tS4:11'} are equivalent because if $\psi(x)$
transforms as $\bs{1}$ then $\psi'(x)=i\psi(x)$ transforms as $\bs{1}'$ (notice
that the field must be complex).
This same reasoning leads to the equivalence of $\bs{3}$ and $\bs{3}'$ when we go
from $S_4$ to $\tS_4$.

Let us see what happens to the representation $\bs{2}$ of $S_4$, which is
equivalent to the $S_4\to S_3$ homomorphism.
To preserve the structure of $S_4\otimes\aver{\cp}$ for which $\cp$ commutes with
the elements of $S_4$, it is important to consider real representations of $\bs{2}$.
We adopt a slightly different version of Ref.\,\cite[Eq.(640), in second reference]{hagedorn},
\eq{
D_2(S)=\begin{pmatrix} 1 & 0 \cr 0 & -1 \end{pmatrix}\,,\quad
D_2(T)=\begin{pmatrix} 
-\ums{2} & \ums[\sqrt{3}]{2} \cr -\ums[\sqrt{3}]{2} & -\ums{2}
\end{pmatrix}\,.
}
Then $S_4\otimes\aver{\cp}$ in this representation is generated by $D_2(S),D_2(T)$
and $\cp$ acting as in \eqref{cp:def}.
The subgroup $\tS_4$ would be generated by
\eq{
D_2(\tS)=\begin{pmatrix} 1 & 0 \cr 0 & -1 \end{pmatrix}\cdot\cp\,,\quad
D_2(T)=\begin{pmatrix} 
-\ums{2} & \ums[\sqrt{3}]{2} \cr -\ums[\sqrt{3}]{2} & -\ums{2}
\end{pmatrix}\,.
}

However, it is usually more convenient to work with the complex basis where $T$ is
diagonal. We change basis to
\eq{
\label{tS4:2}
D_2'(\tS)=\begin{pmatrix} 1 & 0 \cr 0 & 1 \end{pmatrix}\cdot\cp\,,\quad
D_2'(T)=\begin{pmatrix} \om & 0 \cr 0 & \om^2 \end{pmatrix}\,,
}
where
\eq{
D_2'(T)=X^\dag D_2(T)X\,,
}
with the basis-change matrix
\eq{
X=\frac{1}{\sqrt{2}}\begin{pmatrix}1 & 1\cr i & -i\end{pmatrix}\,.
}
Now, $D_2'(\tS)$ in \eqref{tS4:2} differs from 
\eq{
X^\dag\begin{pmatrix} 1 & 0 \cr 0 & -1 \end{pmatrix}X\cdot\cp
}
because $X$ is complex and complex basis change acts differently for CP
transformations. The correct transformation is 
\eq{
D_2'(\tS)=X^\dag\begin{pmatrix} 1 & 0 \cr 0 & -1 \end{pmatrix}X^*\cdot\cp\,,
}
which leads to \eqref{tS4:2}.

Equation \eqref{tS4:2} defines the representation of $\tS_4$, derived from $\bs{2}$
of $S_4$. Since both transformations which generate $\tS_4$ do not mix the first
and second components, they are essentially one dimensional (complex). They
correspond to the representations which we denoted by $\bs{1}_\om$ and
$\bs{1}_{\om^2}$, corresponding to the action of \eqref{tS4:2} to the first and
second components, respectively.
Explicitly, for a fermion field $\psi(x)$ (chiral or not), we have
\eqali{
\bs{1}_\om:&~
\psi(x)\stackrel{\tS}{\longrightarrow} C\psi^*(\hx)\,,\quad
\psi(x)\stackrel{T}{\longrightarrow} \omega\,\psi(x)\,,\cr
\bs{1}_{\om^2}:&~
\psi(x)\stackrel{\tS}{\longrightarrow} C\psi^*(\hx)\,,\quad
\psi(x)\stackrel{T}{\longrightarrow} \omega^2\psi(x)\,.
}
If could ignore gauge quantum numbers, the representation $\bs{1}_\om$ and
$\bs{1}_{\om^2}$ would be equivalent because if $\psi(x)\sim \bs{1}_\om$, then
$C\psi^*(x)\sim \bs{1}_{\om^2}$. Its real representation space is two dimensional.
In particular, $\bs{1}$ and $\bs{1}'$ would correspond to CP-even and CP-odd
combinations of fields which have no definite transformation properties under the
gauge groups.
It is important to emphasize that if we were considering the whole
$S_4\otimes\aver{\cp}$, the representation $\bs{2}$ would remain two dimensional
(complex).


\end{document}